\documentclass[conference]{IEEEtran}

\usepackage{cite}
\usepackage{amsmath,amssymb,amsfonts}
\usepackage{algorithmic}
\usepackage{graphicx}
\usepackage{textcomp}
\usepackage{xcolor}
\usepackage[ruled,linesnumberedhidden]{algorithm2e} 
\usepackage{flushend}

\usepackage{caption}
\usepackage{subcaption}
\usepackage{textpos}
\usepackage{framed}

\usepackage[textsize=tiny]{todonotes}
\usepackage{comment}

\def\BibTeX{{\rm B\kern-.05em{\sc i\kern-.025em b}\kern-.08em
    T\kern-.1667em\lower.7ex\hbox{E}\kern-.125emX}}

\usepackage[hyphens]{url}
\usepackage{hyperref}
\usepackage[hyphenbreaks]{breakurl}

\begin{document}

\title{Advancing Blockchain-based Federated Learning through Verifiable Off-chain Computations} 

\author{\IEEEauthorblockN{Jonathan Heiss, Elias Grünewald, Stefan Tai}
\IEEEauthorblockA{\textit{Information Systems Engineering} \\
\textit{TU Berlin}\\
Berlin, Germany \\
\{jh,eg,st\}@ise.tu-berlin.de}
\and
\IEEEauthorblockN{Nikolas Haimerl}
\IEEEauthorblockA{\textit{Distributed Systems Group} \\
\textit{TU Wien}\\
Wien, Austria \\
s01452766@student.tuwien.ac.at}
\and
\IEEEauthorblockN{Stefan Schulte}
\IEEEauthorblockA{\textit{Christian Doppler Laboratory}\\
\textit{for Blockchain Technologies} \\
\textit{for the Internet of Things, TU Hamburg}\\
Hamburg, Germany \\
stefan.schulte@tuhh.de}
}

\maketitle

\begin{textblock*}{\textwidth}(0cm,-6.9cm) 
\begin{center}
    \large{--~\emph{Extended Version (Preprint)}~--}
\end{center}
\end{textblock*}

\begin{textblock*}{\textwidth}(0cm,17.5cm) 
\begin{center}
\begin{framed}
    \textit{Preprint (2022-06-23) before final copy-editing. Extended version of a peer-reviewed paper to appear in the\\ Proceedings of the \textbf{5\textsuperscript{th} IEEE International Conference on Blockchain (Blockchain 2022)}.}
\end{framed}
\end{center}
\end{textblock*}


\begin{abstract}
Federated learning may be subject to both global aggregation attacks and distributed poisoning attacks. Blockchain technology along with incentive and penalty mechanisms have been suggested to counter these. In this paper, we explore verifiable off-chain computations (VOC) using zero-knowledge proofs (ZKP) as an alternative to incentive and penalty mechanisms in blockchain-based federated learning. In our solution, learning nodes, in addition to their computational duties, act as off-chain provers submitting proofs to attest computational correctness of parameters that can be verified on the blockchain. We demonstrate and evaluate our solution through a health monitoring use case and proof-of-concept implementation leveraging the ZoKrates language and tools for smart contract-based on-chain model management. Our research introduces verifiability of correctness of learning processes, thus advancing blockchain-based federated learning significantly through VOC.   

\end{abstract}
\begin{IEEEkeywords}
federated learning, blockchain, verifiable off-chain computations, zero knowledge proofs, privacy
\end{IEEEkeywords}

\section{Introduction}
\label{sec:introduction}
Machine learning is these days seen as an indispensable feature in many different fields, for instance smart cities~\cite{ullah20} or smart healthcare~\cite{li21}. Despite the potential benefits of machine learning, privacy concerns are very often named as a major hindrance for its wide-spread adoption. This is especially the case if personal data needs to be processed~\cite{FedLearn_Healthcare_2018}. In conventional settings, machine learning is conducted in a centralized style, very often making use of cloud-based computational resources for the training of a machine learning model. This may lead to a large communication overhead, if data from a large number of distributed data sources, e.g., in the Internet of Things (IoT), needs to be provided to the cloud for the learning tasks. 

In order to overcome these issues, \emph{federated learning} offers a different approach to machine learning. 
Instead of the traditional centralized setting, federated learning distributes the model training among nodes, which are often located at the edge of the network and therefore close to the data sources~\cite{DBLP:journals/corr/abs-1902-01046,9084352}. Hence, the training data does not have to be shared with any centralized node. Instead, only the learned model is shared with some aggregation entity. Apart from the apparent improvement with regard to data privacy due to the lack of sharing of raw training data, this also decreases the communication overhead significantly. 

Because of its distributed nature, federated learning may however suffer from malicious contributors, i.e., distributed learners which provide an inaccurate or even fake local model to the aggregator~\cite{fang20,ma20}. Such distributed \emph{poisoning attacks} may decrease the integrity of the resulting machine learning model severely. In addition, the aggregator may itself attack the integrity of the resulting model, if the aggregation process is not done in a transparent way. For instance, the aggregator could skew the resulting model by weighting some local models higher than others. This is also known as a \emph{global aggregation attack}. Therefore, it is necessary to provide mechanisms which impede the occurrence or consequences of attacks both by the decentralized learners and the aggregators.

Several recent studies have proposed to apply blockchain technologies for this, i.e., to do the model aggregation on-chain~\cite{li_chen_liu_huang_zheng_yan_2020,lu_huang_dai_maharjan_zhang_2020,Node-aware_weightinhMethods_2019,BLADE_Performance_Resources_2020,trustworthyFedLearn_2021,articleKim}, and to offer incentive and penalty mechanisms which provide a stimulus to the decentralized learners to deliver verified local models~\cite{privacygradient,deepChain,FLChain,two-layered_architect}. One common issue with these solutions are the occurring costs for verifying the local models: Especially if making use of a public blockchain, carrying out heavyweight computations in smart contracts may lead to (very) high gas costs~\cite{eberhardt_tai_2018}, which diminishes the number of application scenarios for which blockchain-based federated learning is useful.

In this paper, we propose an alternative system architecture for trustworthy federated learning, leveraging smart contracts to aggregate global learning models, and zero-knowledge proofs to make the computational correctness of the local learning processes verifiable on the blockchain. For this, we model and implement federated learning using blockchain technology along with verifiable off-chain computations (VOC). Our contributions can be summarized as follows:

\begin{itemize}
    \item We propose and implement a system architecture for federated learning that leverages blockchain technology to provide decentralized, tamper-resistant, and globally verifiable management of global learning parameters and VOC to enable verifying the computational correctness of local learning processes on the blockchain. 
	\item We instantiate this architecture in a scenario where sensitive data serve as inputs to the federated learning system illustrating the feasibility of our approach. 
	\item We evaluate our findings with regard to learning performance and computational costs. Moreover, we extensively discuss the architecture with regard to alternative deployments and possible extensions. 
\end{itemize}
The remainder of this paper is structured as follows:
First, we summarize preliminaries on federated learning, blockchains, and ZKPs in Section~\ref{sec:background}. Afterwards, we introduce our general system design (Section~\ref{sec:systemDesign}), present a specific showcase scenario from the field of smart healthcare in Section~\ref{sec:application}, and describe the system's implementation (Section~\ref{sec:Implementation}). We apply the presented scenario in the evaluation (Section~\ref{sec:Evaluation}), and discuss the results (Section~\ref{sec:Discussion}). Finally, we assess the related work (Section~\ref{sec:relatedWork}) and conclude our paper (Section~\ref{sec:Conclusion}).

\section{Background}
\label{sec:background}
Our research combines the two previously separated concepts of federated learning and VOC. 

\subsection{Federated Learning}
Federated learning provides an architectural approach to apply machine learning models to a distributed setting where many nodes jointly train a shared global model in an efficient and privacy-preserving manner~\cite{fedLearn_originalPaper_2016}. 

A learning model typically consists of learning parameters, e.g., weights and biases. 
In Artificial Neural Networks (ANNs), the learning parameters are contained by the network's hidden layers and applied to the input data during forward propagation to determine the learning output, e.g., a prediction class.  
In ANNs for supervised learning, this learning output is compared to predetermined truth labels during backpropagation to determine the learning error. 
Based on that error, the learning parameters are re-calculated and then applied to the model's hidden layers for the next learning cycle~\cite{bebis1994feed}.

In federated learning, learning parameters are computed as local models by distributed learning nodes on local data. 
Each participating node provides only its model to an aggregator that combines all received models into a global model which represents the collective learning insights of the learning network. 
This global model is then locally applied by the learning nodes in the next training cycle. 
Thereby, each learning node benefits from the training executed at other nodes without revealing potentially confidential inputs~\cite{fedLearn_originalPaper_2016,yang2019federated}.

While federated learning has originally been proposed as centralized architecture~\cite{yang2019federated} where the global model is managed by a single aggregator, decentralized architectures have also been proposed where the management of the global model is jointly executed by the learning nodes, e.g., ~\cite{li_chen_liu_huang_zheng_yan_2020,lu_huang_dai_maharjan_zhang_2020,Node-aware_weightinhMethods_2019,BLADE_Performance_Resources_2020,trustworthyFedLearn_2021,articleKim,privacygradient,deepChain,FLChain,two-layered_architect}.

\subsection{Verifiable Off-chain Computations}
\label{sub:voc}
Blockchain  technology enables mutually distrusting stakeholders to jointly maintain state in a trustless manner, that means, no single party needs to be trusted, but instead the system design itself provides trust guarantees. 
It does this through both, a cryptographically secured, append-only data structure and an incentive-driven~\cite{wood2014ethereum} or Byzantine~\cite{PBFT_castro_2002} consensus protocol.
While the design of blockchains enables tamper-resistance and public verifiability in distributed environments and, thereby, represents a good fit for managing global parameters in federated learning settings, it also causes weak scalability and leaking privacy. 
One conceptual approach to mitigate these deficiencies are VOC.

In VOC~\cite{off-chaining_models_heiss}, computation that would originally be executed on-chain is outsourced to an arbitrary off-chain node. 
In addition to the computational result, the off-chain node generates a proof that attests to the computation's correctness and can be verified on the blockchain. 
Thereby, VOC enables on-chain verification of computations on confidential data without revealing it on-chain. 
Furthermore, it mitigates scalability limitations since an arbitrary large computation can be executed non-redundantly by a scalable off-chain node and only a fixed-size proof smaller than the computation itself is verified as part of the costly consensus protocol. 
Thereby, extending blockchain-based federated learning with VOC enables on-chain verifiability of off-chain learning processes. 

As enabler for VOC, \emph{ZoKrates}~\cite{eberhardt_tai_2018} has been proposed, a technology that provides a language and toolbox for zkSNARKs-based VOC. 
zkSNARKs (zero-Knowledge Succinct Non-interactive Argument of Knowledge) represent a specific zero-knowledge protocol that distinguishes through non-interactivity and succinct proof sizes. 
ZoKrates hides the peculiarities of zkSNARKs and provides a convenient, developer-friendly means for implementing and deploying both, the off-chain proving program and the on-chain verification smart contract for the Ethereum blockchain~\cite{wood2014ethereum}.
In this paper, we leverage ZoKrates to implement the local learning models as a proof-of-concept of our system design.

\section{System Design}
\label{sec:systemDesign}
Addressing prevalent security issues in federated learning systems exposed by the local model training and the global model management, we propose a system design that strives towards the following objectives: 

\begin{itemize}
    \item \emph{Tamper-resistance}: No party can maliciously manipulate a learning model, i.e., neither the local nor the global model.
    \item \emph{Global verifiability}: The computation of the local and the global model must be verifiable by all participating nodes.  
\end{itemize}
To achieve these objectives, our system architecture has two distinguishing features. 
First, the global model is stored on the blockchain and managed by smart contracts. This enables public verifiability and tamper resistance through system guarantees of the underlying blockchain. 

Second, the local model is implemented as a realization of VOC as described in Section~\ref{sub:voc}. This extends public verifiability and tamper resistance towards computations executed off-chain on the local model while preserving the privacy guarantees of federated learning. 
We thereby address the initially stated problem of non-trusted learning nodes that weaken the global model's integrity by deliberately or accidentally corrupting local computations that return false inputs for the global model updates.

\subsection{System Overview}
\label{sub:overview}
\begin{figure}[t] %
	\centering
	\includegraphics[width=1\columnwidth]{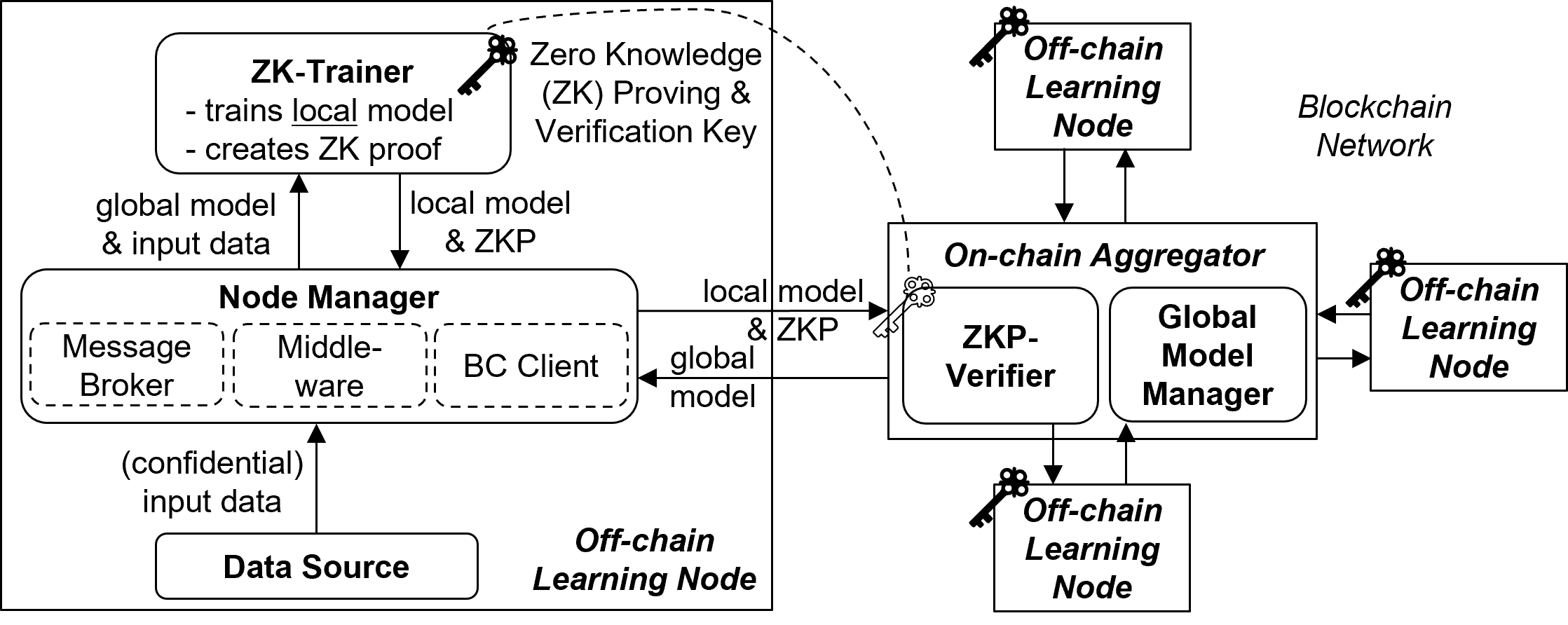}
	\caption{Overview}
	\label{fig:overview}
\end{figure}

The system, as depicted in Figure~\ref{fig:overview},
can be described from the two federated learning perspectives: the local perspective comprises components running on individual \emph{off-chain learning nodes} and the global perspective considers the joint management of the global model through all learning nodes redundantly executing the \emph{on-chain aggregator}.

\subsubsection{Off-chain Learning Node}
On the off-chain learning node, the \emph{data source} creates the data that is input to the local training process and may contain confidential information. 

The \emph{node manager} handles the interaction between the data source, zk-trainer, and the on-chain aggregator. 
For this, it implements a \emph{message broker} to accept messages from the data source which in turn are provided through the middleware to the zk-trainer. 
Furthermore, it implements a \emph{blockchain client} to handle bi-directional communication with the on-chain aggregator. The local model updates are transformed into blockchain transactions and forwarded to the on-chain aggregator. 
In turn, the blockchain client reads global model updates from the on-chain aggregator and provides them to the zk-trainer through the middleware. 
The \emph{middleware} manages inputs to and outputs of the zk-trainer. 

The \emph{zk-trainer} implements a learning algorithm as a zero-knowledge computation to train the local model using both, inputs from the data source and the iteratively updated global model. 
In addition to the updated local model, the zk-trainer produces a ZKP that attests to the computational correctness of the local training processes. 
To create the ZKP, the zk-trainer uses a dedicated proving key that is part of an asymmetric key pair. 
ZKPs are verified using the corresponding verification key which is part of the on-chain aggregator. 

\subsubsection{On-chain Aggregator}
The on-chain aggregator is implemented as smart contracts that are redundantly executed by the participating learning nodes on a blockchain infrastructure. 
The on-chain aggregator has two components: one for verifying ZKPs and one for managing the global model.   

The \emph{ZKP-verifier} uses the verification key that is part of the asymmetric key pair to verify ZKPs attached to local model updates. 
Thereby, it validates that the updates have been computed correctly. 
Since the local off-chain computation is the same on all learning nodes, the same proving key is used for all off-chain proof constructions. 
Consequently, the on-chain aggregator can use a single verification key to verify that the updated local models are output of the expected off-chain computation. 

Once a ZKP is verified, the \emph{global model manager} can aggregate the local model updates into the global model. 
Serving as a single endpoint, the on-chain aggregator receives these updates from all participating learning nodes and, thereby, maintains a single source of truth throughout the nodes of the blockchain network. 
As a central challenge, the on-chain aggregator must, thereby, guarantee fairness, i.e., only one update is applied per node in each cycle, and liveness, i.e., a new cycle is initiated eventually. 
Therefore, the on-chain aggregator keeps a list of all registered participants which, for simplicity, we assume to be predetermined. Addressing fairness, the on-chain aggregator checks that the learning nodes only submit one update per cycle at max. 
To guarantee liveness, the on-chain aggregator initiates a new cycle after a predefined period, determined based on the block-based timestamp. 
This is done even if it has not yet received an update from each registered learning node.


\subsection{Workflow}
\begin{figure}[ht] %
    \centering
    \includegraphics[width=1\columnwidth]{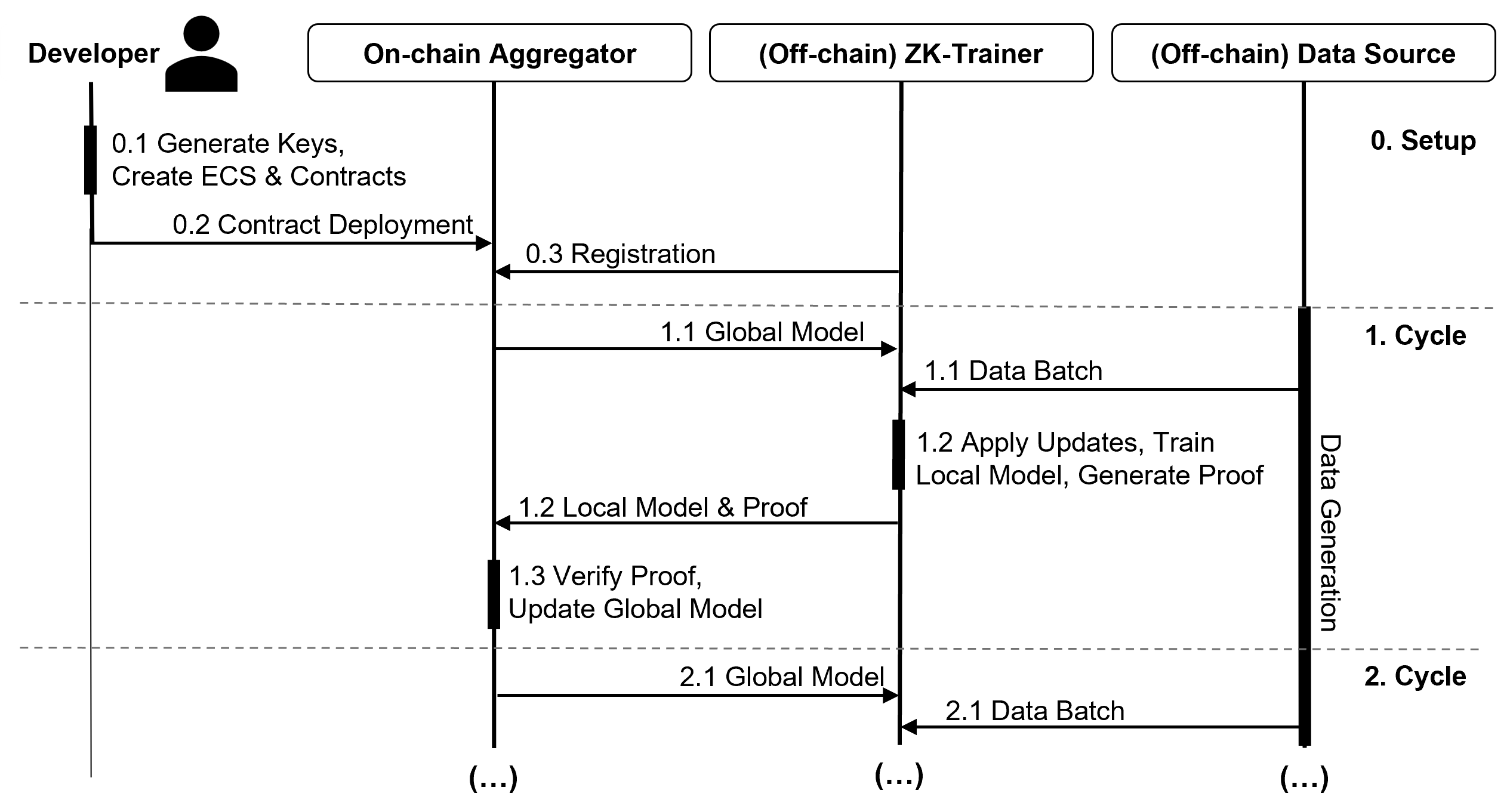}
    \caption{Workflow}
    \label{fig:procedure}
\end{figure}

As depicted in Figure~\ref{fig:procedure}, the system workflow comprises an initial one-time \emph{setup} and repeated \emph{updating cycles}. 
In the workflow, four parties are involved: the developer who is exclusively involved in the setup phase, the on-chain aggregator, the zk-trainer, and the data source. For simplicity, in Figure~\ref{fig:procedure}, the on-chain aggregator is represented as a single component and the node manager is omitted since it only manages interactions. 

\label{sub:workflow}

\emph{Setup}: 
During the one-time setup, the developer creates the zk-trainer as the off-chain prover and the on-chain aggregator as the on-chain verifier. 
In this paper, we assume that this VOC instantiation is realized with ZoKrates and, hence, follows the setup flow described in~\cite{eberhardt_tai_2018}: The developer defines the training algorithm in the ZoKrates high-level language, compiles it into an executable constraint system (ECS), and, based on the ECS, generates the proving-verification key pair. 
The zk-trainer requires the proving key and the ECS, whereas the on-chain aggregator holds the verification key and implements the ZKP verification logic. 
The implemented on-chain aggregator contracts for global model management and ZKP verification are then deployed to the blockchain.
Once deployed, learning nodes can be registered with their blockchain account address for participation. 

\emph{Updating Cycles}:
At the beginning of a cycle, the zk-trainer reads the latest global model from the on-chain aggregator and applies the contained parameters, e.g. weights and biases, to the local model. 
It receives the inputs for the training as a data batch of fixed size and data format from the data source. 
Based on the global model parameters and the input data, the zk-trainer executes the training as a zero-knowledge computation in two steps. 
First, it executes the ECS which returns a witness, i.e., an input-specific variable assignment of the ECS. Second, the ZKP is generated using this witness and the proving key. 
The resulting ZKP and the updated local model are then sent to the on-chain aggregator as a blockchain transcation. 
On receiving the transaction, the smart contract first verifies the proof with the verification key and, if successful, updates the global model with the received local model parameters.


\section{Application}
\label{sec:application}
The proposed system architecture is applicable to various federated learning contexts. 
Without mitigating this generality, in this section, we introduce an application context where the guarantee of the model’s integrity is of particular interest and, hence, the verifiability of the local training and the global model management of paramount importance. 
Then, we derive a suitable learning model that fits to the characteristics of the application context and our proposed system architecture.

\subsection{Use Case}
\label{sub:useCase}
Remote, \emph{in-home health tracking} through wearables is particularly relevant if an individual’s health condition restricts its mobility such that routine examinations at the doctor or hospitals cannot be fulfilled anymore~\cite{FedHome_wu_2020,Hierarical_fedLearn_healthcare_2021}. 
In such contexts, ``smart'' or learning-based health applications that monitor an individual’s health condition through wearables and predict worsening health conditions can help saving lives, e.g., by informing healthcare services or relatives on time.  
Such applications often provide classification problems, e.g., predicting health states (good, fair, serious, critical) from measurements collected by wearables.

Federated learning has been applied for in-home health monitoring to improve the learning models of such smart health applications through a joint training of a global learning model by connected households~\cite{FedHome_wu_2020,Hierarical_fedLearn_healthcare_2021}. 
Since personal health data is highly confidential and requires special protection as required by privacy regulations such as the EU’s GDPR~\footnote{\url{https://gdpr.eu}} or the United States’ HIPAA~\footnote{\url{https://www.hhs.gov/hipaa/for-professionals/privacy/index.html}}, federated learning can help comply with these regulations and keep this data private. 
Furthermore, the integrity of the global learning model is of utmost important since false predictions of the health condition can have dramatic consequences, e.g., lead to death. 
Therefore, guarantees of global verifiability and tamper resistance have high priority in such application contexts. This makes our system proposal particularly suitable for such a scenario.

If an in-home health monitoring system as proposed in~\cite{FedHome_wu_2020} or~\cite{Hierarical_fedLearn_healthcare_2021} is realized with our system architecture, each household represents a learning node. 
The data source is represented by a wearable, e.g., a smartwatch, which is worn by the patient and continuously generates health data. 
The smartwatch transmits the data to an in-home workstation which hosts all other components of the learning node and the on-chain aggregator's smart contracts. 
Therefore, it also provides the networking interface to other connected household and healthcare services. 

\subsection{Learning Model}
\label{sub:learningModel}
\begin{figure}[ht] %
	\centering
	\includegraphics[width=1\columnwidth]{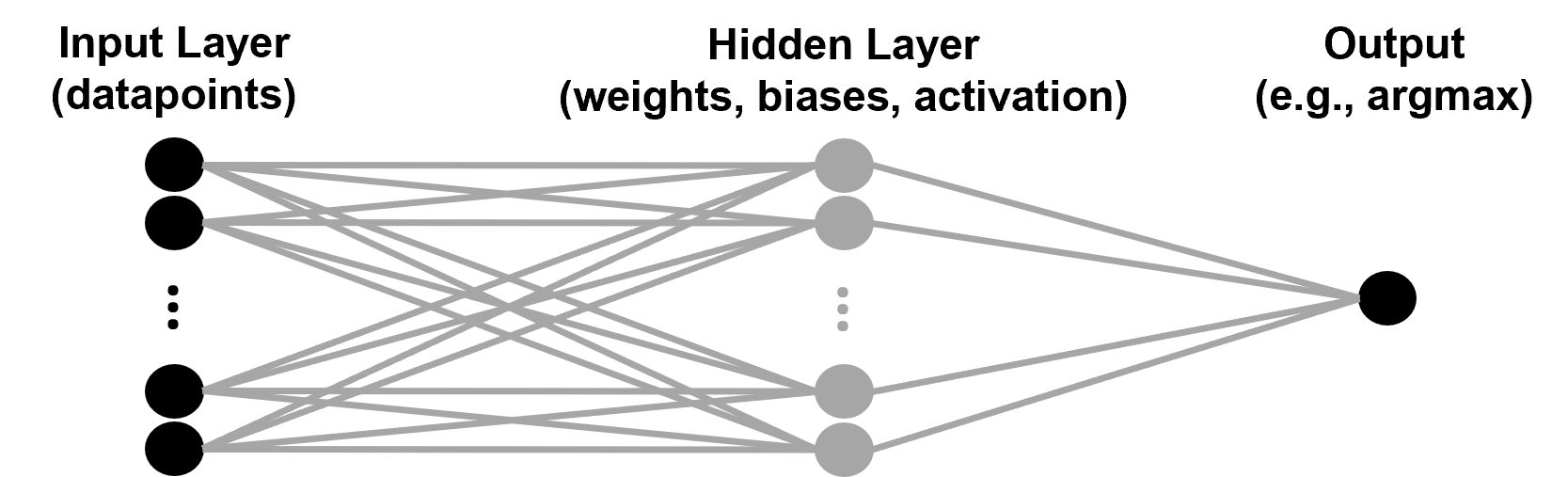}
	\caption{Feedforward Neural Network with One Hidden Layer}
	\label{fig:learningModel}
\end{figure}

In order to define a learning model for the described smart healthcare use case, we select and design an ANN with respect to the following aspects: 

\begin{itemize}
    \item Learning problem: The selected use case provides a classification problem where sensor data inputs must be correctly assigned to a predefined set of prediction classes. These classes define the status of the user. For the learning model design, aspects of the inputs, i.e., sensor measurements, and the outputs, i.e., prediction classes, must be taken into consideration. 
    \item Execution environment: The construction of a verifiable ZKP adds a considerable computational overhead to the local training that depends on the input data size and the training's computational complexity~\cite{Jacob_Diss_2021}. Consequently, a simple and efficient learning model is desirable.
\end{itemize}
In this respect, we select a simple \emph{feedforward neural network}~\cite{bebis1994feed} with one hidden layer as depicted in Figure~\ref{fig:learningModel}.

A feedforward neural network is an ANN where information always moves in one direction. 
As defined in the following, the depicted feedforward neural network consists of three layers: The input layer which represents a n-sized vector $X$ of datapoints, a hidden layer which consists of a m-sized vector $Y$ of biases and a m-n-sized matrix of weights $W$, and the output layer which, in this case, represents a single prediction class. 

\[
    X=
    \begin{bmatrix}
    x_1 \\
     \vdots \\
    x_n
    \end{bmatrix},
       B=
    \begin{bmatrix}
    b_1 \\
       \vdots \\
    b_m
    \end{bmatrix},
    W=\left[
\begin{array}{ccc}
   w_{11} & \cdots & w_{1m} \\
   \vdots & \ddots & \vdots \\
   w_{n1} & \cdots & m_{nm}
\end{array}
\right]
\]
The network is iteratively trained through cycles of forward propagation and backpropagation. Learning insights generated in one cycle are applied in the next cycle to improve the weights and biases of the hidden layer. 

During \emph{forward propagation}, datapoints of the input layer are fed through the hidden layer of the network to determine a prediction class.
On the hidden layer, the inputs are multiplied with the weights, the biases are added, and the activation function $\sigma$ is applied which projects the calculated values to the set of predetermined prediction classes:

\[    
    \hat{Y}=\sigma(XW + B)
         \quad\text{or}\quad 
    \hat{y}_j=\sigma(b_j+\sum_i x_iw_{ij})
\]
From the $n$-sized output vector $\hat{Y}$, an aggregation function determines the most suitable class as the output, in our case, an argmax function which simply selects the highest value. 

During \emph{backpropagation}, the prediction’s quality is evaluated and the neural network’s weights and biases are updated. 
Using a loss function, the difference between the prediction and the truth labels is determined. In our case, we realize the loss function with the mean squared error:

\[
    L=\frac{1}{n}\sum_i^n(y_i-\hat{y}_i)^2 
    \text{,}\quad 
    \frac{\partial L  }{\partial \hat{y}_j}=\frac{1}{n}(y_j-\hat{y}_j)
\]
Using the loss function’s derivatives and the learning rate $\alpha$, the gradient is calculated, and the new weights $W^{\prime}$ and biases $B^{\prime}$ are determined:
\[
    W^{\prime}:=W-\alpha \frac{\partial L  }{\partial \hat{y}_j}   \frac{\partial \hat{y}_j  }{\partial \sigma_{j}}\frac{\partial \sigma_j  }{\partial w_{ij}}
             \quad\text{,}\quad
    B^{\prime}:=B-\alpha \frac{\partial L  }{\partial \hat{y}_j}   \frac{\partial \hat{y}_j  }{\partial \sigma_{j}}\frac{\partial \sigma_j  }{\partial b_{j}}.
\]
The learning rate determines the impact that the gradient imposes on the update of the current weights and is defined as a hyperparameter prior to the training. 
The updated weights and biases are applied to the hidden layer and leveraged in the next cycle.

Importantly, while we apply the discussed feedforward neural network in the implementation at hand, arbitrary learning models could be applied. 

\section{Implementation}
\label{sec:Implementation}
To demonstrate technical feasibility of our proposal, we implement the presented system as a proof-of-concept (PoC) for the previously introduced application context and realize the described feedforward neural network as a ZoKrates-enabled VOC\footnote{\url{https://github.com/NikolasHaimerl/Advancing-Blockchain-Based-Federated-Learning-Through-Verifiable-Off-Chain-Computations.git}}).

\subsection{Off-chain Learning Node}
In our PoC, the zk-trainer and the node manager represent the core components of the off-chain learning node. 

\subsubsection{ZK-Trainer}
As most distinguishing aspect of our PoC, the previously described feedforward neural network is implemented using the ZoKrates\footnote{\url{https://github.com/Zokrates/ZoKrates}} high-level language and executed as a zkSNARKs-based computation. 
For this, we leverage ZoKrates CLI commands with their current default settings, i.e., the Groth16 proving scheme~\cite{groth16_jensGroth_2016} and the alt\_bn128 (Barreto-Naehrig) curve.

The program takes the current batch of input data, the learning rate as hyperparameter, and the latest global model as input arguments.
Based on the arguments, it executes the matrix calculations on the hidden layer and applies the argmax function to determine the prediction. 
The loss function is implemented as the mean squared error function and its derivative as described in Section~\ref{sub:learningModel}.
The ZoKrates program is compiled into an ECS as described in Section~\ref{sub:workflow} and, then, accessible through its main function.
It returns ZKP as well as the new weights and biases representing the updated local model.

While providing Turing-completeness, the ZoKrates programming language comes with some constraints that impact our implementation of the learning algorithm, noteworthily with regards to two aspects:
First, in ZoKrates only unsigned integers can be used as data types. Consequently, float numbers contained in the input data are scaled to enable integer-based operations.
Furthermore, for every value, a corresponding Boolean is passed to the program which indicates whether the value is positive or negative. Arithmetic operations are rewritten to account for possible over- and underflows and return the signs of the resulting values. 
    
Second, in ZoKrates, the size and the type of the input arguments must be determined at compile time. Consequently, the number of datapoints of the input layer cannot be dynamically changed. This also impacts the implementation of exponential activation functions where exponents are dynamically calculated based on the input data. Hence, in our PoC implementation, only linear relationships between the features and truth labels are predicted. 

\subsubsection{Node Manager}
The node manager's \emph{message broker} is implemented with RabbitMQ\footnote{\url{https://www.rabbitmq.com/}}, a 
broker that enables simple communication with low power consumption, suitable for IoT settings.
It supports multiple data sources and, for each, provides a dedicated message queue. This does not only enable off-chain learning nodes with multiple data sources, but also helps to realize the testbed for our experiments described in Section~\ref{sec:Evaluation}.

Published messages are consumed by the \emph{middleware} which is implemented in Python. 
To construct the input arguments for the zk-trainer, it obtains the input data from the message broker and the updated global model from the on-chain aggregator through the blockchain client. Fixed input parameters, such as the learning rate, are retrieved from local storage.
The middleware executes instances of the zk-trainer by calling the ZoKrates CLI: First, the witness is computed based on the ECS with the \texttt{compute-witness} command which takes all input arguments. Second, the proof is generated with the \texttt{generate-proof} command which takes the witness and the proving key as arguments.
The intermediate witness file is written to and read from the local file system.
Outputs are provided to the on-chain aggregator through the blockchain client. 

The \emph{blockchain client} is also implemented in Python and relies on the web3.py library\footnote{\url{https://github.com/ethereum/web3.py}} to communicate with the on-chain aggregator.
It uses the blockchain account address of the local learning node to transmit the ZKP and the local model as a blockcain transaction to the on-chain aggregator.

\subsection{On-chain Aggregator}
The on-chain aggregator is implemented in Solidity\footnote{\url{https://github.com/ethereum/solidity}} and separates the functionality into two Ethereum~\cite{wood2014ethereum} smart contracts: the learning and the verifier contract. 
Ethereum is chosen, since it is a very widely used smart contract-based blockchain that allows for public and private deployment of smart contracts.

\subsubsection{Learning Contract}
The learning contract provides the updating mechanism for the global model and, as the common endpoint for the off-chain learning nodes, implements the management of the updating cycles introduced in Section~\ref{sec:systemDesign}.

Analogue to the local model, the global model consists of the weight-vector and the bias-matrix as described in Section~\ref{sub:learningModel}.
The updating mechanism requires the smart contract to store two versions: 
One that is only updated at the end of the cycle and is consumable by off-chain learning nodes. 
And a temporary one that is used to aggregate the updated local models of the participating learning nodes using a simple moving average function. 
At the end of an updating cycle, the temporary model replaces the other. 

To enforce that only one update per learning node and cycle is applied to the global model, the learning contract keeps a list of all updates made during a cycle by using the respective account addresses of the learning nodes. 
Furthermore, it determines the beginning of a new cycle through a time interval that is set by the constructor and based on the block-based timestamp.

\subsubsection{Verifier Contract}
The verifier contract is called from the learning contract before an update is applied to the global model.
It implements the ZKP verification logic in the \emph{verifyTx()} function and treats the verification key as hard-coded parameter. 
The contract is automatically generated as a Solidity smart contract by means of the ZoKrates CLI. 


\section{Evaluation}
\label{sec:Evaluation}

\begin{figure*}
\centering
    \begin{minipage}[b]{.48\textwidth}
        \includegraphics[width=1.0\columnwidth]{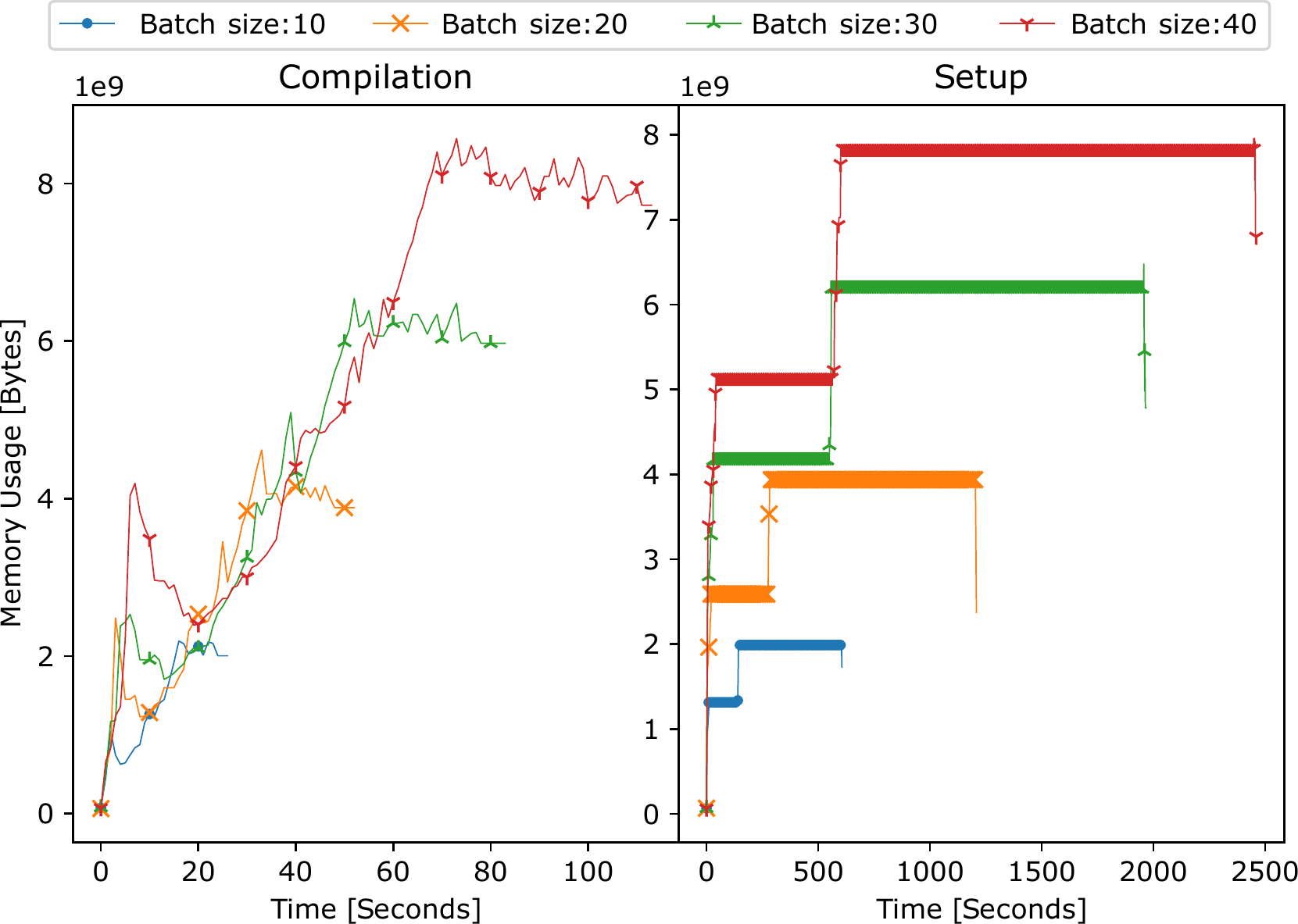}
        \caption{Memory requirements of compilation and setup with different batch sizes.}
        \label{fig:memoryCS}
    \end{minipage}\qquad
    \begin{minipage}[b]{.48\textwidth}
     \includegraphics[width=1.0\columnwidth]{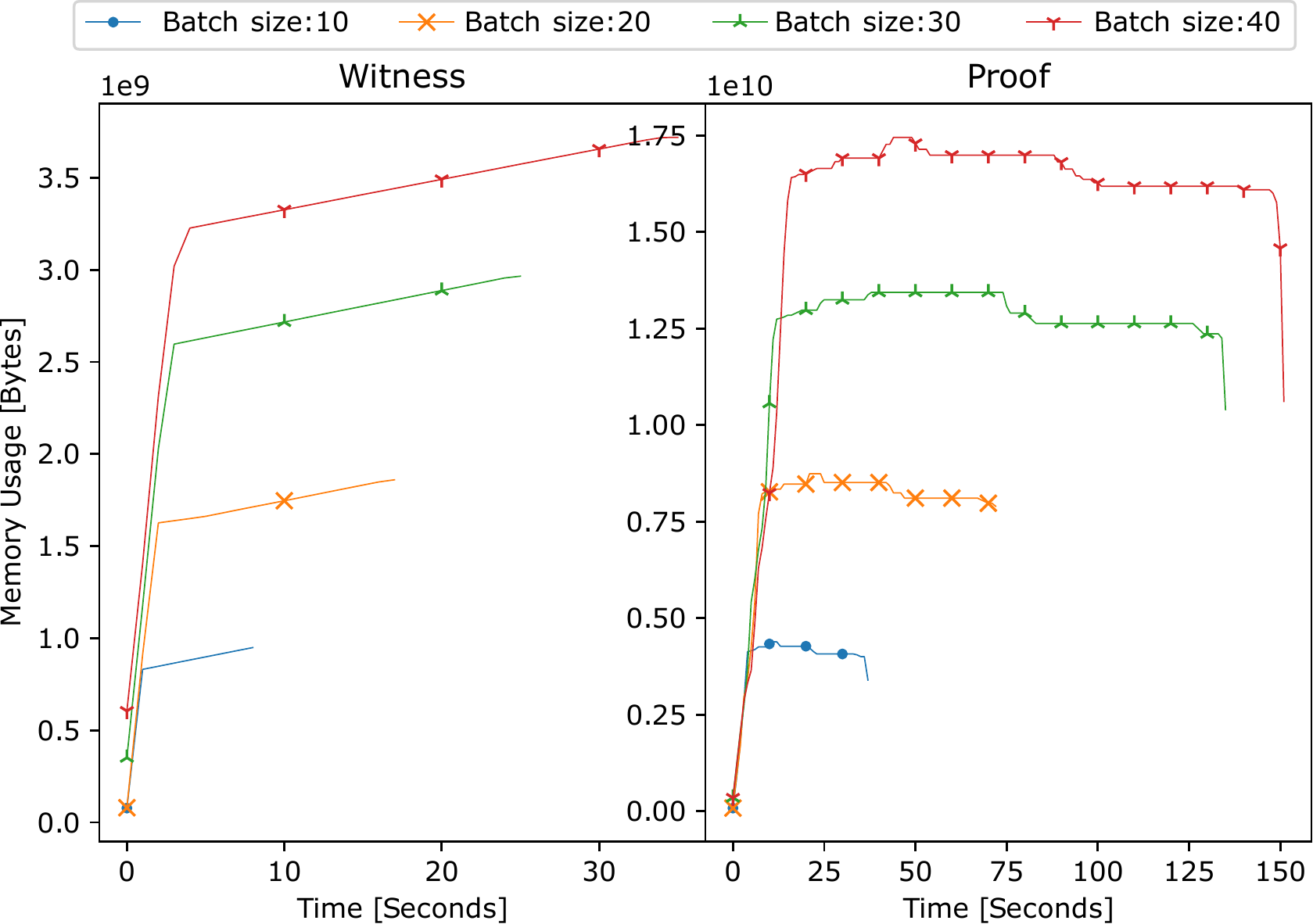}
        \caption{Memory requirements of computing witnesses and the proof with different batch sizes.}
        \label{fig:memoryPW}
    \end{minipage}
\end{figure*}


To evaluate the system, we deploy the PoC implementation in a test environment and analyze its behavior during experimentation.
Therefore, we first select a dataset that fits to our application context, then describe our experimental setup, and finally discuss results obtained from the experiments with regards to computational and learning performance.

\subsection{Dataset}
\label{sub:dataset}
To represent the described use case, we select a public dataset that is closely related to the described use case in that it is built upon sensor data generated from wearables and provides a classification problem.
The dataset is available under the UCI Repository for Daily and Sports Activities Data Set\footnote{~\url{https://archive.ics.uci.edu/ml/datasets/daily+and+sports+activities\#}} and has been generated by eight subjects (four females and four males between the ages 20 and 30) who executed activities for a duration of five minutes each.
To adopt the dataset to our use case and make it suitable for our proposed system, we apply two major changes:

First, we reduce the feature space. In total, the dataset comprises 45 features which originate from measurements in different spatial directions generated by different sensor devices attached to different body parts. 
To adopt the dataset to our use case where a single smartwatch is used as data source, we reduce the dataset to measurements of one sensor device resulting in nine features which determines the size of the neural network's input vector.

Second, we reduce the set of prediction classes. The dataset comprises 19 different activities, i.e., prediction classes. 
To reduce complexity and align the data with our use case and learning model, we combined activities that characterize through similar measurements, e.g., ascending and descending stairs, or exercise in a stepper and on a cross trainer. 
This results in a total of six prediction classes which determines the size of the hidden layer's biases vector and the weight matrix.

\subsection{Experimental Setup}
Aligned with our use case, we spawn eight in-home health monitoring systems 
that act as off-chain learning nodes and are managed by a smart contract-based on-chain aggregator. The on-chain aggregator runs on a virtual blockchain network which is instantiated with Ganache\footnote{\url{https://github.com/trufflesuite/ganache}}. 
Each household is represented by one data source, one ZoKrates-based zk-trainer, and one blockchain account.
To simulate the different data sources, we implement a Python-based \emph{workload generator} that retrieves the dataset described in Section~\ref{sub:dataset} from the filesystem and splits it into eight subsets each representing a data source. 
From the subsets, the workload generator publishes data batches from each data source to the corresponding queue provided by the message broker.

Given that our node manager implementation is capable of managing multiple data sources and multiple zk-trainers, it is not replicated for each household but instead, in our experimental setup, manages multiple households instances. 
As such, it employs eight zk-trainers which run in parallel during a cycle.
The resulting local models are provided to the on-chain aggregator through the Ganache blockchain client which provides the models of each household through the corresponding account address.

To simulate this environment and conduct the experiments, we use a single machine equipped with an AMD Ryzen 3950X@3.5~GHz CPU with 16~cores, memory capacity of 128~GB@3600 MHz (DDR4), and a solid-state-drive with approximately 3500~MB/s read and write capabilities. 
The experiments are executed three times and the results represent the average of the measurements gathered in each run. In addition, we show the standard deviations, where meaningful.

\subsection{Computational Costs}
To gain insights about the computational performance of our system, we observe off- and on-chain computations during the one-time setup and the updating cycles discussed in Section~\ref{sec:systemDesign}.
We measure the execution time~[sec] and memory consumption~[Bytes] for off-chain computations and transaction costs~[Gas\footnote{Gas is an Ethereum-specific metric for measuring transaction complexity. 
}] for on-chain computation. 


To understand the system behavior for varying workloads, we conducted the experiments with four different batch sizes 
comprising 10/20/30/40 datapoints respectively per cycle (cf. Table~\ref{tab:eval_numbers}). 

The \emph{setup phase} comprises two off-chain computations: the compilation of the ZoKrates program and the key generation (in Figure~\ref{fig:memoryCS} referred to as setup following the ZoKrates command name), and the on-chain aggregator deployment as on-chain computations. 
As depicted in Figure~\ref{fig:memoryCS}, execution time and memory consumption behave rather linearly to the increasing batch sizes. 
While during the key generation (setup) the memory consumption is rather static, it grows over the execution time during compilation. 
The latter may be caused by an increasing memory space allocated for the constraint system which is built during compilation. The initial one-time deployment cost of the on-chain aggregator comprises 0.00624~Gas for the \textit{ZKP-verifier} and 0.0023~Gas for the \textit{global model manager} smart contracts. 

During the \emph{operational phase}, in each cycle the witness generation and the proof construction are executed as off-chain computations whereas proof verification and the model update are executed on-chain. 
As depicted in Figure~\ref{fig:memoryPW}, memory size and execution time grows rather linearly with increasing batch sizes. Clearly, proof construction is more expensive than witness generation. The transaction costs for verification range between approximately 0.0024~Gas (batch size: 10) and 0.0025~Gas (batch size: 40)
per cycle. However, compared to the one-time setup, repeating computations in the operational phase require considerably less resources, take less time, and are therefore evaluated as practically feasible.




\begin{table}[]
    \centering
    \caption{Average execution time and memory consumption for varying batch sizes.}
    \begin{tabular}{l|cccc}
         Batch size [\#] & 10 & 20 & 30 & 40  \\
         \hline
         \hline
         Compile Time [sec] & 27.79 & 56.55 & 93.01 & 127.46 \\
         \textit{STD [sec]} & 0.02 & 0.87 & 3.06 & 5.70\\
         Compile Memory [GB] & 2.26 & 4.62 & 6.82 & 9.31 \\
         \textit{STD [MB]} & 80.5 & 18.9 & 18.0 & 56.3 \\
         \hline
         Setup Time [sec] & 651.35 & 1290.40 & 2107.88 & 2711.53 \\
         \textit{STD [sec]} & 2.00 & 1.20 & 7.33 & 56.87 \\
         Setup Memory [GB] & 2.02 & 4.12 & 6.68 & 8.28 \\
         \textit{STD [MB]} & 23.82 & 0.00 & 157.51 & 114.405\\
         \hline
         Witness Time [sec] & 8.90 & 19.21 & 27.43 & 37.76 \\
         \textit{STD [sec]} & 0.50 & 1.50 & 0.50 & 0.50\\
         Witness Memory [GB] & 0.95 & 1.88 & 2.96 & 3.72 \\
         \textit{STD [MB]} & 2.45 & 0.00 & 0.00 & 0.00\\
         \hline
         Proof Time [sec] & 39.55 & 78.25 & 145.50 & 167.56 \\
         \textit{STD [sec]} & 0.03 & 1.08 & 1.05 & 4.34\\
         Proof Memory [GB] & 4.39 & 8.74 & 13.43 & 17.45 \\
         \textit{STD [MB]} & 0.00 & 0.00 & 0.00 & 0.00\\
    \end{tabular}
    \label{tab:eval_numbers}
\end{table}



\subsection{Federated Learning Performance}
\begin{figure}[!t]
 \scalebox{0.45}{\input{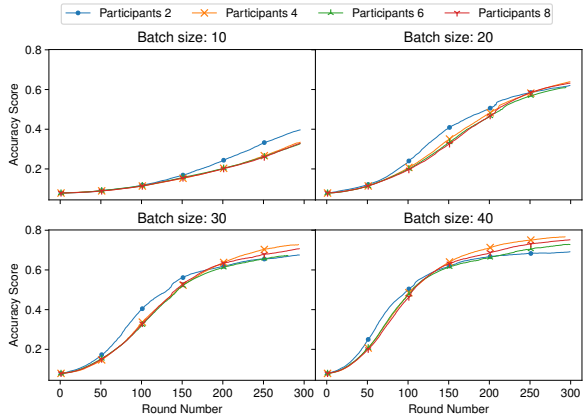}}
    \caption{Accuracy scores for different numbers of participants and various batch sizes.}
    \label{fig:roundScoresParts}
\end{figure}

In addition to the computational performance, insights about the performance of the applied learning model are essential to evaluate its applicability. 
Therefore, we measure the model’s accuracy and how it behaves with a varying number of participating off-chain learning nodes (2, 4, 6, 8). 
Again, experiments are conducted with the four different batch sizes (10, 20, 30, 40) with a total number of 300 training cycles each. Different input data batches are also randomly chosen from the data set to avoid over-fitting of the learning model. Results are depicted in Figure~\ref{fig:roundScoresParts}.

As expected, the more datapoints are used per cycle, the faster the model's output gains accuracy. 
While the improvement curves for batch sizes 20, 30 and 40 all have their inflection point within 300 updating rounds, this is not yet the case for batch size 10. With higher batch sizes, a better accuracy is reached even more early.
Our model converges towards a 0.79 accuracy score with batch size 40, which is considered a very satisfactory result with respect to the simple feed-forward neural network applied in our proof-of-concept architecture due to the limitations imposed by the ZoKrates language. With a more sophisticated model -- which itself was deliberately not subject of our optimizations -- the accuracy is expected to be even higher.

Moreover, while all runs seem to converge towards the same maximum accuracy performance eventually, a discrepancy is noticeable between the number of participants involved. Contrary to what was first expected, the test runs with only 2 learning nodes started off better than the tests with 4, 6 or 8 participating nodes in all variations of the batch size. After investigation, this was caused by the diversity of the data assigned to the nodes. Naturally, some nodes are better suited to train a global model since their data are by chance closer to the general median than others. Therefore, we consider this a minor irregularity in the dataset which would not occur in real-world use cases with more participants.

Lastly, with more participants, a diminishing performance improvement can be observed. While more participants may result in a slower start of the training improvement, the convergence accuracy score is the highest with the most participants. This is to be expected as the experiment’s data were sampled from all nodes.

\section{Discussion}
\label{sec:Discussion}
The system's evaluation underlines its technical feasibility in an exemplary application context but also reveals limitations.
Addressing these, we first discuss different deployment scenarios and then present a system extension for end-to-end integrity between data source and on-chain aggregator.

\subsection{Deployment Scenarios}
\label{sub:deploymentscenarios}
The system's resource consumption, i.e., for local off-chain training and on-chain model management, can impact its deployability. 

Computational overhead added through the ZKP generation as well as the connectivity-effort added through the blockchain-based consensus protocol can present an impediment for the system's deployment, particularly in a constrained environment. 
If the learning problem is non-critical, as in various non-healthcare use cases, and the resources are limited, e.g., on mobile devices, the added overhead of our system may be disproportional. 
Otherwise, however, if more powerful hardware is available and the learning problem is critical, such as in in-home health monitoring use cases, the resource consumption of our proposed system is bearable and its guarantees required. 

On the blockchain, the per-cycle transaction costs depend on the total cost of an individual update and the number of participating learning nodes. 
Transactional costs, however, have a different impact on the participants and the system with regards to the applied blockchain type. 
A deployment to permissionless blockchains, such as Ethereum~\cite{wood2014ethereum} relies on weaker trust assumptions and enables public accessibility and verifiability beyond the defined set of learning nodes through third parties.
However, given incentive-based consensus protocols, e.g., Proof-of-Work (PoW) or Proof-of-Stake (PoS) that rely on cryptocurrencies and require monetary transaction fees, the costs for running the system may quickly become too expensive. 
This problem is less severe for a deployment on a permissioned blockchain, such as Hyperledger Fabric~\cite{HLFabric_originalPaper_2018}.
Here, all nodes are known in advance and can be held accountable such that the consensus protocol can be realized more efficient and without monetary incentives. 
While in this case operational costs may be lower, open accessibility and public, third-party verifiability are restricted.

\subsection{End-to-end Integrity}
\label{sub:end-to-end}
While our proposed model protects against attacks on the local training and the global model management, the off-chain learning nodes still expose an attack vector through the data source: If false data is injected, the local model is corrupted regardless the application of VOC. 

To better understand this attack scenario in our proposed system, we distinguish between two cases: 
In the first case, the data source is controlled by the learning node as depicted in Figure~\ref{fig:overview}. 
By protecting against attacks during local model training, our trust model assumes the learning node to be distrusted in that it either accidentally or deliberately corrupts the local training. 
However, if the data source reveals another attack vector, we must adjust the trust assumptions for the learning nodes towards trustworthy data provisioning.

In the second case, the data source is controlled by an external third party. 
If we assume that this party provides the data in a trustworthy way, the learning node could still corrupt the data before feeding it into the local training without being noticed:
Even if the data is signed by the data source, after the training, the signature cannot be verified by the on-chain aggregator without the input data.

While for the former case no practical solution exists that does not add further trust assumptions, for the latter case end-to-end integrity can be provided by applying the trustworthy pre-processing model presented in~\cite{heiss_trustworthypreprocessing_ICSOC2021}. 
Trustworthy pre-processing enables a smart contract to verify authenticity and integrity of data that has been signed by a trusted data source but has been pre-processed by a distrusted intermediary, e.g., an oracle~\cite{trustworthyOnchaining_heiss} or the off-chain learning node. In~\cite{heiss_trustworthypreprocessing_ICSOC2021}, it has been shown that the pre-processing model can, among others, be realized by applying zkSNARKs.
Consequently, it represents a viable extension to our system to remove trust assumptions from the learning node and to enable end-to-end-integrity between the data source and the on-chain aggregator.

\section{Related Work}
\label{sec:relatedWork}
In this section, we discuss related work around the two attack vectors that we mitigate with our system proposal, i.e., the local model training and the global model management.

Federated learning with a single, centralized aggregator reveals a single point of failure and a surface for malicious model manipulation through a non-trustworthy aggregation node. 
Addressing these limitations of a centralized architecture, in various works, more decentralized setups have been proposed, with blockchains representing one possible technological enabler 
~\cite{li_chen_liu_huang_zheng_yan_2020,lu_huang_dai_maharjan_zhang_2020,Node-aware_weightinhMethods_2019,BLADE_Performance_Resources_2020,trustworthyFedLearn_2021,articleKim}. 
A blockchain-based setup, however, also introduces undesirable overheads that can quickly stress resource capacities in constrained environments. In this regard, literature provides approaches for more efficient federated learning implementations as general approaches~\cite{li_chen_liu_huang_zheng_yan_2020} or simply by selecting permissioned blockchains that implement more lightweight consensus protocols by design, e.g., \cite{FabricFL}.
While our proof-of-concept has not been optimized for storage and computation, our proposal is blockchain agnostic which allows the developer to decide for a deployment according to requirements of the use case at hand.

The other attack vector is exposed by the local training which, especially in distributed systems can leverage for poisoning attacks by malicious learning nodes. 
Addressing this attack in blockchain-based architectures, incentive schemes have been proposed that reward honestly acting nodes while penalizing dishonest nodes in turn~\cite{privacygradient,deepChain,FLChain,two-layered_architect}. 
However, it cannot be enforced in a decentralized manner without a cryptocurrency which makes permissioned blockchains inapplicable. 
Furthermore, it is assumed that learning nodes act as \emph{homini oeconomici}, steadily trying to increase their utility. 
Compared to our approach, incentive schemes rely on stronger assumptions and highly rely on cryptocurrency-enabled blockchains which restricts their applicability. 

As another approach, the local training and the global model management can be executed by Trusted Execution Environments (TEE) which guarantee for computational correctness and even enable non-revealing processing of confidential data~\cite{TEE_Oblivious_2016,TEE_Multi-Source_2018,TEE_secureBoost_Leakage_2020,TEE_trainingIntegrity_2020,TEE_PPFL_2021}. 
TEEs provide hardware-level protection against external access to data in use, i.e., during runtime, and allow third parties to verify the authenticity of messages using hardware-level asymmetric encryption. 
While TEEs represent an alternative to ZKPs to realize off-chain computations as described in~\cite{off-chaining_models_heiss}, to date, they have not been used in blockchain-based federated learning setups. 
Compared with ZKPs, TEEs have weaker security guarantees and rely on a trusted third party, i.e., the hardware manufacturer, which is not required for ZKPs. 
However, they provide better performance and have a smaller computational overhead as, for example, shown in~\cite{heiss_trustworthypreprocessing_ICSOC2021}.

\section{Conclusion}
\label{sec:Conclusion}
We presented a model and first system architecture for blockchain-based federated learning with verifiable off-chain computations. 
We instantiated the architecture in an e-health scenario and evaluated and discussed our implementation and findings also regarding learning performance and computational costs. 
We can conclude that on-chain verifiability of correctness of off-chain learning processes not only is feasible in specific deployments and learning scenarios, but presents an important if not significant direction to federated learning and privacy-enhanced decentralized applications in general.


In our future work, we want to seize on aspects discussed in Section~\ref{sub:end-to-end} by further investigating mechanisms to decrease the costs for verifying federated learning results, e.g., by off-chaining parts of the necessary storage, and to remove yet existing trust assumptions of the learning nodes, e.g., through trustworthy pre-processing~\cite{heiss_trustworthypreprocessing_ICSOC2021}. 
In addition, we want to take a look at approaches to scale the solution to handle larger problem sizes.


\bibliographystyle{IEEEtran}
\bibliography{references}

\end{document}